\documentclass[useAMS,usenatbib]{mn2e}

\usepackage{epsfig}
\usepackage{amsmath}
\usepackage{natbib}
\usepackage{times}
\usepackage{amsfonts}
\usepackage{aas_macros}

\title[Precision Prediction of Log Power Spectrum]{Precision Prediction of the Log Power Spectrum}
\author[A. Repp and I. Szapudi]{A. Repp$^{1}$ and I. Szapudi$^{1}$\\
$^{1}$Institute for Astronomy, 2680 Woodlawn Dr., Honolulu, Hawaii 96822, USA}

\begin{document}
\date{\today}				

\pagerange{\pageref{firstpage}--\pageref{lastpage}} \pubyear{0000}

\maketitle

\label{firstpage}

\begin{abstract}
At translinear scales, the log power spectrum captures significantly more cosmological information
than the standard power spectrum. At high wavenumbers $k$,
the Fisher information in the standard power spectrum $P(k)$ fails to increase in proportion to $k$
in part due to correlations between large- and small-scale modes. As a result, $P(k)$ suffers from an
information plateau on these translinear scales, so that analysis with the standard power spectrum
cannot access the information contained in these small-scale modes. The log power spectrum
$P_A(k)$, on the other hand, captures the majority of this otherwise lost information. 
Until now there has been no means of predicting the amplitude of the log power spectrum apart from
cataloging the results of simulations. We here present a cosmology-independent prescription for the
log power spectrum; this prescription displays accuracy comparable to that of \citet{Smith_et_al},
over a range of redshifts and smoothing scales, and for wavenumbers up to $1.5h$ Mpc$^{-1}$. 
\end{abstract}


\label{firstpage}

\section{Introduction}
\label{sec-intro}
Three-dimensional galaxy surveys can potentially yield significant gains in our
understanding of cosmological parameters. Realizing this potential, however, depends on our
ability to extract the cosmological information inherent in translinear modes.

The power spectrum $P(k)$ is the standard means of elucidating such information (e.g.,
\citealp{Peebles1980,BaumgartFry1991}); and if one is analyzing a Gaussian field, this statistic
exhausts the field's information\footnote{We use the term ``information'' as a shorthand
for the Fisher information content \citep{Fisher1925} of the probability density function of the
matter fluctuation field (e.g., \citealp{Tegmark1997}).}.
Given
that inflation models typically predict a high
degree of Gaussianity for primordial fluctuations (e.g., \citealp{Bardeen1986, BondEfstathiou1987});
given that the Cosmic Microwave Background (CMB) indeed displays a high degree of Gaussianity; 
and given that the evolution of the power spectrum at small wavenumbers is essentially
linear, it follows that the matter power spectrum is an effective summary statistic for capturing
the large-scale information in three-dimensional surveys.

On smaller (translinear) scales, however, gravitational amplification of the original fluctuations
alters the dark matter field to a distinctly non-Gaussian distribution \citep{FryPeebles1978,
Sharp1984, Szapudi1992, Bouchet1993, Gaztanaga1994}. As a result of this nonlinear evolution,
the distribution develops a long non-Gaussian tail; this tail produces large cosmic variance,
since stochastic occurrence of massive clusters disproportionately affects
the power spectrum on such scales \citep{Neyrinck2006}.

This increase in cosmic variance involves a corresponding decrease in the information
content of the power spectrum. One aspect of this information loss is the rise of correlations
between the values of $P(k)$, so that extending the
survey to smaller scales only modestly increases the Fisher information in $P(k)$. Through this and other
mechanisms, a significant amount of information escapes from the power spectrum, producing
a marked information plateau at such wavenumbers
\citep{RimesHamilton2005, NeyrinckSzapudi2007, LeePen2008, Carron2011, CarronNeyrinck2012,
Wolk2013}.

\citet{Repp2015} have shown that for amplitude-like parameters, the power spectrum on translinear scales
can contain an order of magnitude less information than it would for a Gaussian field.
Since survey forecasts typically assume Gaussianity, they can thus overestimate
a survey's effectiveness (and hence its effective volume) by a factor of two or more.

Higher order statistics (e.g., $N$-point correlation functions) can access some \citep{Szapudi2009}
but not all of \citep{CarronNeyrinck2012, CarronSzapudi2013} this information; these
statistics also suffer from difficulties in calculation and interpretation.

The log transformation, on the other hand, is particularly attractive as a means of accessing this information
\citep{Neyrinck2006, NSS09} in that it emerges naturally under the assumption
of linear growth of peculiar velocities \citep{ColesJones1991}. Despite its simple analytic form,
\citet{SzapudiKaiser2003} have shown that this transformation is equivalent to an infinite-order loop
perturbation theory. In addition, simulations have indicated that the shape of the log power
spectrum tracks that of the linear power spectrum up to high wavenumbers \citep{NSS09}, thus
reversing the effects of nonlinear evolution.

\citet{CarronSzapudi2013} investigate observables which can extract
all of the cosmological information inherent in a field. Statistics constructed from such observables
(``sufficient statistics'') would, if tractable, be the optimal statistics to use in analyzing matter
distributions. In particular, they consider the observable $A$ produced by applying the log transformation to
the overdensity field $\delta = \rho/\overline{\rho} - 1$:
\begin{equation}
A = \ln (1 + \delta).
\end{equation}
\citeauthor{CarronSzapudi2013} show that the observable $A$ differs from the true optimal observable
by a negligible amount, as long as the power spectrum slope is reasonably close to $-1$. Thus,
despite the fact that the Universe's matter distribution is only approximately lognormal, the
log overdensity $A$ is essentially a sufficient statistic for these fields.

One could argue, of course, that in some sense the ``information'' in the fields $\delta$
and $A$ is identical, given the existence of a known invertible mapping between them. However,
what \citeauthor{CarronSzapudi2013} (among others)
show is that at high wavenumbers, the Fisher information in the log power spectrum $P_A(k)$ is
comparable to the information in the field---whereas the power spectrum $P(k)$ contains
significantly less information, as would indeed any combination of higher-order statistics
of the $\delta$ field.

The log density fluctuation $A$, together with its power spectrum $P_A(k)$, is thus an ideal means
of completely extracting cosmological
information on translinear scales. However, the primary barrier to the use of $P_A(k)$ has been
the lack of a theory predicting its value for various cosmologies. This lack leaves simulations as the
primary---and computationally expensive---means of exploring $P_A(k)$.

In this letter we present a simple prescription for predicting the log power spectrum $P_A(k)$.
This prescription contains only one phenomenological parameter, and it provides accuracy
comparable to that of \citet{Smith_et_al} as refined by \citet{Takahashi2012}. For the
remainder of this letter we use ST to denote this standard prescription of \citeauthor{Smith_et_al}
and \citeauthor{Takahashi2012}

We organize this letter as follows: Section~\ref{sec:method} outlines the process by which we derive
our basic prescription. Section~\ref{sec:accuracy} quantifies the accuracy of our prescription; it also
demonstrates that including a slope modulation parameter substantially increases the accuracy.
Section~\ref{sec:discussion} discusses potential future refinement to and application of the
prescription; and Section~\ref{sec:concl} summarizes this work.

\section{Method}
\label{sec:method}
The Millennium Simulation \citep{Springel2005} includes enough dark matter particles to prevent
$1/\overline{n}$ discreteness from affecting its power spectrum. Since $A$ is a sufficient statistic
for such continuous fields, this simulation is ideal for investigating $P_A(k)$.
We therefore obtained\footnote{http://gavo.mpa-garching.mpg.de/Millennium/}
snapshots corresponding to $z = 0, 0.1, 0.5, 1.0, 1.5,$ and 2.1.
In the original Millennium Simulation cosmology (which we assume until Section~\ref{sec:cosmo}),
each of the $256^3$ cells has sides of length 1.95 Mpc$/h$. In addition, we smoothed the data to cells twice, four times, and
eight times this size, yielding smoothing scales up to 15.6 Mpc$/h$. Using the prescription of
\citet{Jing2005} to compensate for pixel window and aliasing effects, we obtained both
$P(k)$ and $P_A(k)$ for these six redshifts and these four smoothing scales.

\begin{figure}
    \leavevmode\epsfxsize=9cm\epsfbox{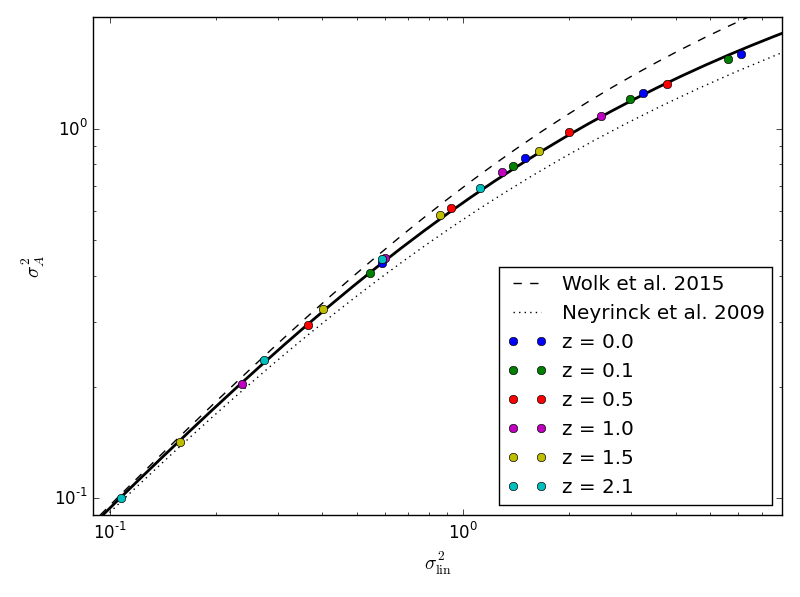}
    \caption{Variance of log overdensity versus linear variance, at multiple redshifts and smoothing
    scales. The solid curve shows the logarithmic fit of Equation~\ref{eq:sigA} with $\mu=0.73$.
    We also show, for comparison, the relationships implied by the bias approximations of \citet{NSS09}
    and \citet{WCS2015}.}
\label{fig:var_plot}
\end{figure}

Although \citet{NSS09} indicate that the shape of $P_A(k)$ matches that of the
linear power spectrum $P_\mathrm{lin}(k)$, the former spectrum is biased with respect to the latter, so
that $P_A(k) = b_A^2 P_\mathrm{lin}(k)$. Various approximations for this
bias exist (\citealp{NSS09, WCS2015}---see Fig.~\ref{fig:var_plot}), but they 
are not sufficiently accurate to facilitate the use of the log power spectrum in constraining
cosmological parameters.

The variance over a field is the integral of its power spectrum;
thus when two power spectra differ by a multiplicative constant, that constant equals
the ratio of the corresponding variances---in
this case, of $\sigma_A^2$ and $\sigma_\mathrm{lin}^2$. To calculate $\sigma_A^2$ for each
snapshot and smoothing scale, we
integrate $P_A(k)$ up to the Nyquist frequency $k_N$. To calculate $\sigma^2_\mathrm{lin}$
we use CAMB (Code for Anisotropies in the Microwave Background\footnote{http://camb.info/}:
\citealp{CAMB}) to generate the linear power spectrum $P_\mathrm{lin}(k)$ for each snapshot and smoothing
scale;
we then perform the same integral (shown explicitly in Equation~\ref{eq:intPk}). Plotting
$\sigma_A^2$ against $\sigma_\mathrm{lin}^2$, we obtain the tight relationship
in Fig.~\ref{fig:var_plot}.

If the $\delta$ field were precisely lognormal, then the relationship $\sigma_A^2 = \ln(1 + \sigma^2)$
would hold for all variances. Since the lognormal approximation is increasingly accurate at large
scales (and small variances), we would expect, for small values of $\sigma_\mathrm{lin}^2$, that
\begin{equation}
\label{eq:lowend}
\sigma_A^2 \approx \ln(1 + \sigma_\mathrm{lin}^2) \approx \sigma_\mathrm{lin}^2.
\end{equation}

The relationship between the calculated values of $\sigma^2_A$ and $\sigma^2_\mathrm{lin}$
is well-described by a simple logarithmic function of the form
\begin{equation}
\label{eq:sigA}
\sigma_A^2 = \mu \ln\left(1 + \frac{\sigma^2_\mathrm{lin}}{\mu} \right),
\end{equation}
which contains but one free parameter $\mu$. Note that this function has the correct
low-end behavior, reducing to
Equation~\ref{eq:lowend} for low $\sigma^2_\mathrm{lin}$. Using least squares optimization, we obtain a
best fit value of 0.73 for this parameter $\mu$, and we thus use $\mu=0.73$ for the remainder of this letter.

With this relationship we can calculate the bias of $P_A(k)$ with respect to $P_\mathrm{lin}(k)$, so that
\begin{equation}
\label{eq:PA}
P_A(k) = \frac{\sigma^2_A}{\sigma^2_\mathrm{lin}}P_\mathrm{lin}(k).
\end{equation}

In summary, one obtains the log power spectrum $P_A(k)$ by first calculating the linear spectrum $P_\mathrm{lin}(k)$. One then obtains $\sigma_\mathrm{lin}^2$ by integration:
\begin{equation}
\label{eq:intPk}
\sigma^2_\mathrm{lin} = \int_0^{k_N} \frac{dk\, k^2}{2\pi^2} P_\mathrm{lin}(k),
\end{equation}
where the Nyquist frequency $k_N = \pi/\ell$, $\ell$ being the side length of one pixel of the survey volume.
The variance $\sigma^2_A$ of the log field follows from Equation~\ref{eq:sigA}, and the log power spectrum $P_A(k)$ then follows from Equation~\ref{eq:PA}.
\begin{figure*}
    \leavevmode\epsfxsize=18cm\epsfbox{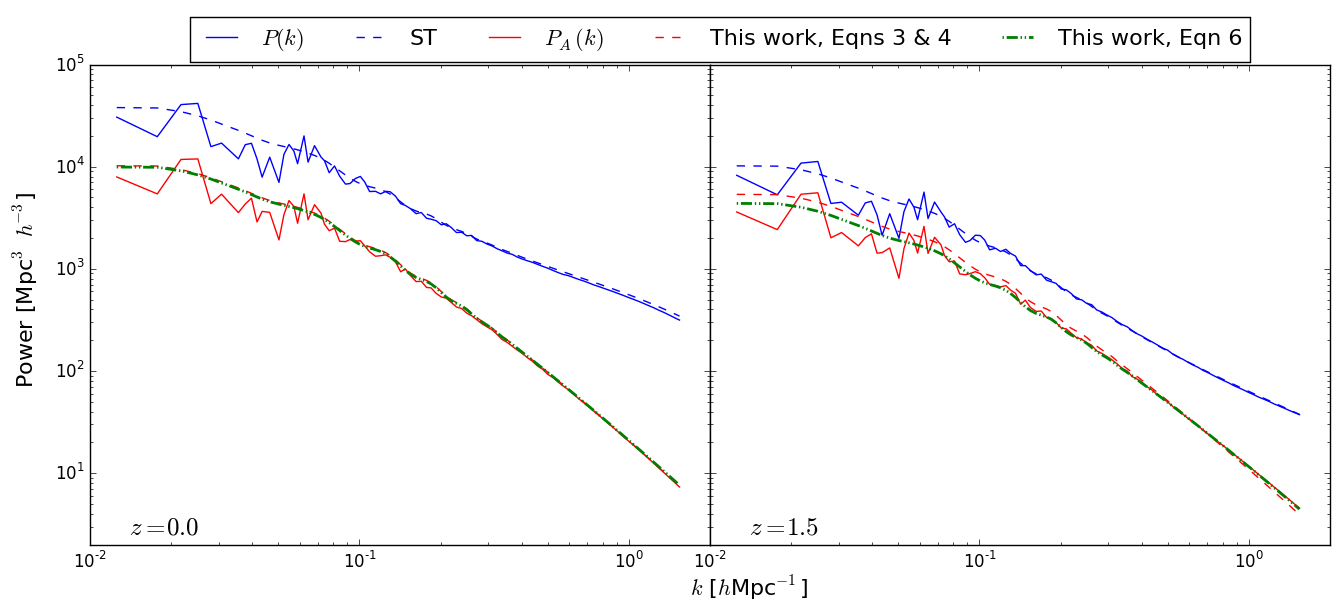}
    \caption{The upper set of curves (colored blue) compares $P(k)$ predicted by ST (dashed)
 with Millennium Simulation results (solid) at $z=0$ and $z=1.5$ with a smoothing scale of
 $\ell =1.95$ Mpc$/h$.
 The lower set of curves (colored red and green) compares
 $P_A(k)$ predicted by Equations~\ref{eq:sigA} and \ref{eq:PA} (dashed) with
 Millennium Simulation results (solid) at the same redshift and smoothing scale. The 
green dashed-dotted curve shows the result of slightly modulating the slope of $P_A(k)$ as described in
Section~\ref{sec:slopemod}. Note that in the left-hand panel, the dashed-dotted green curve obscures
the dashed red curve by coinciding with it.}
\label{fig:example_fit}
\end{figure*}

\begin{figure*}
    \leavevmode\epsfxsize=18cm\epsfbox{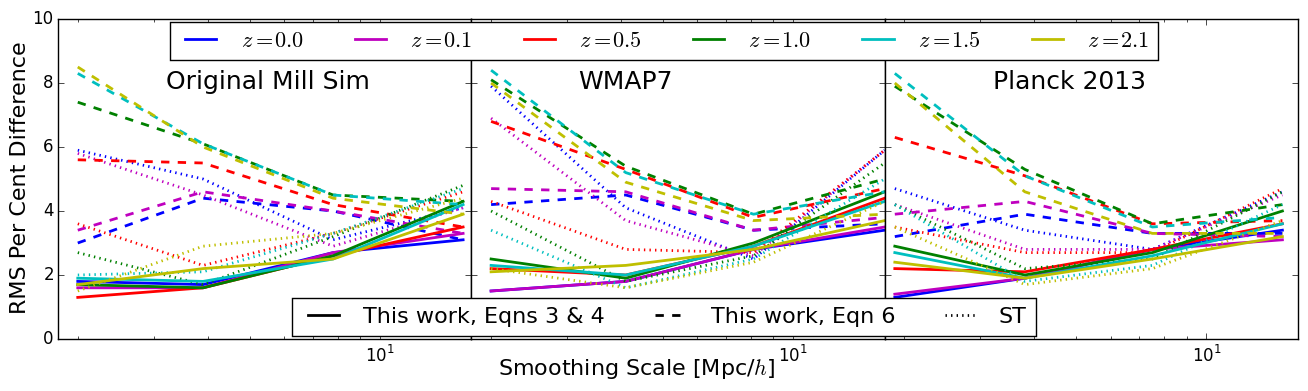}
    \caption{RMS percent differences between various
    prescriptions and the corresponding power spectra, at a variety of redshifts and
    scales ($k > 0.15h$ Mpc$^{-1}$) for three cosmologies. Solid lines show the accuracy of our
    prescription for $P_A(k)$ with the slope modulation of Equation~\ref{eq:slopemod}
    to compensate for residual nonlinearity. Dashed lines denote the accuracy of our prescription 
    (Equations~\ref{eq:sigA} and \ref{eq:PA}) for $P_A(k)$, with no slope modulation. Dotted lines
    denote the accuracy of the ST prescription for the nonlinear $P(k)$. We note that our prescription for
    the log power spectrum $P_A(k)$ is comparable in accuracy to that of \citet{Smith_et_al} and
    \citet{Takahashi2012} for the power spectrum $P(k)$.}
\label{fig:errors}
\end{figure*}

\section{Accuracy}
\label{sec:accuracy}
To quantify the accuracy of our prescription for $P_A(k)$, and to compare it with the standard ST
prescription for $P(k)$, we calculate both power spectra at redshifts
from 0 to 2.1, on smoothing scales from $\ell=1.95$ to 15.6 Mpc$/h$. In performing
this comparison, we exclude
wavenumbers less than $k = 0.15h$ Mpc$^{-1}$ where cosmic variance dominates the inaccuracy.
We then calculate the RMS percent difference of our prescription
and the measured $P_A(k)$, and of the ST prescription and the measured
$P(k)$. The left-hand panel of Fig.~\ref{fig:errors} shows the accuracies, using dotted lines to denote
the results of Equations~\ref{eq:sigA} and \ref{eq:PA}, and solid lines to denote the results of ST.

In general, our prescription has an accuracy of 4--7\%, whereas the ST prescription is
accurate to 3--5\%. Our formula is most accurate at low redshifts; Fig.~\ref{fig:example_fit}
shows an almost perfect correspondence between our prescription and $P_A(k)$ at $z=0$; however,
it also shows that at $z=1.5$ there is a slight but noticeable discrepancy between the slopes
of $P_A(k)$ and $P_\mathrm{lin}(k)$. It is this residual nonlinearity which produces
the increased inaccuracy at higher redshifts.

\subsection{Removing residual non-linearity}
\label{sec:slopemod}
One can thus improve the predictive power of our formula by slightly modulating the high-end
slope. We assume that in the linear regime ($k < 0.15h$ Mpc$^{-1}$) no modulation
is necessary; for higher wavenumbers, we assume that the correction takes the form
$(k/0.15)^\alpha$ for some $\alpha$. We include an additional normalization constant $N$ to insure
that the total variance remains unchanged. Thus, we write
\begin{equation}
\label{eq:slopemod}
P_A(k) = N\,C(k)\,\frac{\sigma^2_A}{\sigma^2_\mathrm{lin}}P_\mathrm{lin}(k),
\end{equation}
where
\begin{equation}
C(k)=\left\{
\begin{array}{ll}
    1	&	\mbox{if $k < 0.15h$ Mpc$^{-1}$} \\
    (k/0.15)^\alpha	& \mbox{if $k \ge 0.15h$ Mpc$^{-1}$}
\end{array}
\right.,
\end{equation}
and
\begin{equation}
\label{eq:N}
N=\frac{\int dk\, k^2 P_\mathrm{lin}(k)}{\int dk\, k^2 C(k) P_\mathrm{lin}(k)}.
\end{equation}

\begin{table}
 % \centering
  \begin{minipage}{8.5cm}
  \caption{Best fit values of slope modulation $\alpha$}
  \label{tab:alpha}
  %\begin{center}
  \begin{tabular}{lcccccc} \hline
 $z$		& 0.0		& 0.1		& 0.5		& 1.0		& 1.5		& 2.1\\
 $\alpha$	& .02		& .04		& .09		& .13		& .14		& .14\\
\hline
 \end{tabular}
 %\end{center}
\end{minipage}
\end{table}

To use this prescription we need only fit the parameter $\alpha$, which is clearly redshift dependent.
We do so using the smallest available smoothing scale ($k_N=1.6h$ Mpc$^{-1}$) for each of our six
redshifts, obtaining the values in Table~\ref{tab:alpha}.

The left-hand panel of Fig.~\ref{fig:errors} shows the RMS percent difference  (for translinear
wavenumbers, as before) between this modulated prescription and the measured $P_A(k)$. One sees that
the overall accuracy of 2--4\% is in general slightly better than that of ST.

\subsection{Cosmology-independence}
\label{sec:cosmo}
The above results assume the original Millennium Simulation cosmology. \citet{AnguloWhite2010}
describe a procedure for rescaling simulations from one cosmology to another. The central element
of their procedure consists of finding the box size and redshift in the target cosmology such that
the linear variance most closely matches the linear variance in the
final snapshot of the original
cosmology. One then compares the relative sizes of the growth function $D(z)$ in the
two cosmologies to match earlier original-cosmology redshifts with the correct target-cosmology
redshifts. Thus a rescaling of the
Millennium Simulation to a different cosmology changes the size of each cell and maps each
redshift to a different snapshot. By definition, the rescaling preserves the linear variance
$\sigma^2_\mathrm{lin}$. However, there is no a priori reason to suppose that it preserves
the relationship between $\sigma^2_\mathrm{lin}$ and $\sigma^2_A$ unless that relationship is
indeed cosmology-independent.

Thus we consider the publicly available rescalings of the Millennium Simulation to the WMAP7 and Planck 2013
cosmologies. We apply our prescription (both the
unmodulated form of Equations~\ref{eq:sigA}
and \ref{eq:PA}, and the slope-modulated form of Equation~\ref{eq:slopemod} and
Table~\ref{tab:alpha}) to these rescalings; when we do so, we find accuracy comparable to that
seen in the original Millennium cosmology. Fig.~\ref{fig:errors} shows that the accuracy of
the unmodulated prescription is 4--7\%, and the accuracy of the modulated prescription is 2--4\%,
in all three cosmologies.

Thus we conclude that our prescription is independent of cosmology, as long as the cosmology
under consideration is reasonably close to the concordance cosmology.

\section{Discussion}
\label{sec:discussion}
We have tested our prescription using the Millennium Simulation only, since most other
simulations are not dense enough to allow one to ignore shot noise effects. For discrete realizations
of low-density fields (where there exists a significant probability of few or no particles in
a cell), the log transformation no longer provides a sufficient statistic. Thus in future work we plan
to generalize our prescription to the power spectrum of $A^*$, which is the optimal
observable analogous to $A$ for discrete fields \citep{CarronSzapudi2014}. This extension will allow us
to further validate our prescription against a variety of cosmological simulations.

However, even absent such testing, the accuracy of this prescription is noteworthy. The matter distribution
of the Universe is only approximately lognormal; nevertheless, it is a striking experimental fact that the
log transform reverses almost all of the effects of nonlinear evolution, leaving a power spectrum virtually
identical in shape to $P_\mathrm{lin}(k)$; furthermore this correspondence holds to wavenumbers past
$k = 1.5h$ Mpc$^{-1}$. The range of applicability of our prescription thus contrasts favorably
with that of perturbation theory, which fails by $k \sim 0.2h$ Mpc$^{-1}$.

In addition, the central formula of our prediction (Equation~\ref{eq:sigA}) requires a 
phenomenological fit of only one parameter, namely, $\mu$. The degree to which the value
of $\mu$ can shed light on the theory of nonlinear evolution (or vice versa) remains to be seen.
However, we currently have no means of quantifying the effect of cosmic variance
on the best-fit value for $\mu$ until we can compare among multiple simulations.
It is curious that this simple relationship (Equation~\ref{eq:sigA}) obtains between
$\sigma_A^2$ and $\sigma_\mathrm{lin}^2$, and not between $\sigma_A^2$ and $\sigma^2$.
However, since through \citet{Smith_et_al} we can obtain the relationship between $\sigma^2$ and
$\sigma^2_\mathrm{lin}$, our relationship means that we can predict $\sigma^2_A$ from
$\sigma$, providing another means of testing our prescription. It turns out that the same is
true for $\langle A \rangle$, which we however do not include in this paper.

Finally, in order to remove residual nonlinearity, we have introduced the slope modulation
parameter $\alpha$. The values listed in Table~\ref{tab:alpha} give consistently accurate predictions
across all cosmologies tested; however, one would expect the residual nonlinearity to be greatest
at $z=0$, which is behavior opposite to what we find for $\alpha$. We do not know whether this
counterintuitive behavior points to possible residual non-Gaussianity from Zel'dovich initial conditions
at these higher
redshifts, and this possibility is another reason for future testing against other simulations. If future work
fails to elucidate its nature, one could nevertheless marginalize over $\alpha$ as a nuisance
parameter.

\section{Conclusion}
\label{sec:concl}

Based on the Millennium Simulation, we have provided a prescription (Equations~\ref{eq:sigA} and
\ref{eq:PA}) for calculating the log power spectrum $P_A(k)$. This prescription is accurate to a few
percent when one includes a slope modulation parameter $\alpha$
(Equations~\ref{eq:slopemod}--\ref{eq:N}), and this accuracy is comparable to (indeed, better than) that
of the \citet{Smith_et_al} prescription for $P(k)$. Our prescription is accurate for redshifts from $z=0$ to at
least 2, on smoothing scales from 2 to 16 Mpc$/h$, and to wavenumbers past
$1.5h$ Mpc$^{-1}$. It is independent of cosmology, as long as one stays reasonably close
to the concordance cosmology.

Previous work has conclusively demonstrated the utility of the log power spectrum in constraining
cosmological parameters. \citet{WCS2015} show that it contains twice as much cosmological
information on parameters such as $\sigma_8$ and $w_0$ (see also \citealp{WCS2015Forecast}).
\citet{Repp2015} show that proper accounting for non-Gaussianity (which use of the log power
spectrum accomplishes) can increase the effective dark energy figure of merit by a factor of three.
Similarly, \citet{Wolk2015} show that this technique can tighten the constraint on neutrino mass by
a factor of three. 

However, to access this information one must precisely predict the log power spectrum
$P_A(k)$. In this work we have prescribed a simple means of doing so, 
and this prescription thus paves the way for a significant increase in the precision of
our cosmological knowledge.

\section*{Acknowledgements}
The Millennium Simulation databases used in this paper and the web application providing online
access to them were constructed as part of the activities of the German Astrophysical Virtual
Observatory (GAVO). IS acknowledges support from National Aeronatics and Space Administration
(NASA) grants NNX12AF83G and NNX10AD53G.

\bibliographystyle{mn2e}
\bibliography{PrecisionPredictionLogPower}

\label{lastpage}
\end{document}